\documentclass{solarphysics}

\usepackage[hyperref,optionalrh]{spr-sola-addons} 
\usepackage{graphicx}        
\usepackage{color}           
\usepackage{breakurl}        
\usepackage{array}
\usepackage{makecell}

\newcommand{\etal}{{\it et al.}}

\newcommand{\sur}[1]{\ensuremath{^{\rm #1}}}
\newcommand{\sous}[1]{\ensuremath{_{\rm #1}}}

\chardef\us=`\_

\begin{document}

\begin{article}
\begin{opening}

\title{Which Photospheric Characteristics are Most Relevant to Active-Region Coronal Mass Ejections?}

%

\author[addressref={aff1,aff2},corref,email={jkonto@noa.gr}]{\inits{I.}\fnm{Ioannis}~\lnm{Kontogiannis}~\orcid{0000-0002-3694-4527}}
\author[addressref={aff2,aff6}]{\inits{M.K.}\fnm{Manolis~K.}~\lnm{Georgoulis}~\orcid{0000-0001-6913-1330}}
\author[addressref=aff3]{\inits{J.A.}\fnm{Jordan~A.}~\lnm{Guerra}~\orcid{0000-0001-8819-9648}}
\author[addressref=aff4]{\inits{S.-H.}\fnm{Sung-Hong}~\lnm{Park}~\orcid{0000-0001-9149-6547}}
\author[addressref=aff5]{\inits{D.S.}\fnm{D.~Shaun}~\lnm{Bloomfield}~\orcid{0000-0002-4183-9895}}

\address[id=aff1]{Leibniz-Institut f\"ur Astrophysik Potsdam (AIP), An der Sternwarte 16, 14482, Potsdam, Germany}
\address[id=aff2]{Research Center for Astronomy and Applied Mathematics (RCAAM) Academy of Athens, 4 Soranou Efesiou Street, Athens, GR-11527, Greece}
\address[id=aff6]{Department of Physics \& Astronomy. Georgia State University, Atlanta, GA 30303, USA}
\address[id=aff3]{Department of Physics, Villanova University, Villanova, PA, USA}
\address[id=aff4]{Institute for Space-Earth Environmental Research (ISEE), Nagoya University, Nagoya, Japan}
\address[id=aff5]{Northumbria University, Newcastle upon Tyne, NE1 8ST, UK}

%
\runningauthor{I.Kontogiannis \etal}
\runningtitle{MPIL Properties and CME Characteristics}

\begin{abstract} 
We investigate the relation between characteristics of coronal mass ejections and parameterizations of the eruptive capability of solar active regions widely used in solar flare prediction schemes. These parameters, some of which are explored for the first time, are properties related to topological features, namely, magnetic polarity inversion lines (MPILs) that indicate large amounts of stored non-potential (\textit{i.e.} free) magnetic energy. We utilize the Space Weather Database of Notifications, Knowledge, Information (DONKI) and the \textit{Large Angle and Spectrometric Coronograph} (LASCO) databases to find flare-associated coronal mass ejections and their kinematic characteristics while properties of MPILs are extracted from \textit{Helioseismic and Magnetic Imager} (HMI) vector magnetic-field observations of active regions to extract the properties of source-region MPILs. The correlation between all properties and the characteristics of CMEs ranges from moderate to very strong. More significant correlations hold particularly for fast CMEs, which are most important in terms of adverse space-weather manifestations. Non-neutralized currents and the length of the main MPIL exhibit significantly stronger correlations than the rest of the properties. This finding supports a causal relationship between coronal mass ejections and non-neutralized electric currents in highly sheared, conspicuous MPILs. In addition, non-neutralized currents and MPIL length carry distinct, independent information as to the eruptive potential of active regions. The combined total amount of non-neutralized electric currents and the length of the main polarity inversion line, therefore, reflect more efficiently than other parameters the eruptive capacity of solar active regions and the CME kinematic characteristics stemming from these regions.
\end{abstract}
%

\end{opening}

\section{Introduction}
     \label{S-Introduction} 
Coronal mass ejections (CMEs) are expulsions of solar coronal plasma into the interplanetary space. They affect dramatically the conditions in the heliosphere and upon reaching geospace they can severely damage space-borne and ground-based infrastructure, disrupting communications, navigation systems, and power grids. CMEs have been increasingly monitored and studied since their discovery in the early 1970s \citep[see][for a historical overview]{gopalswamy16}. CME velocities range between 100--3500\,kms\sur{-1}, with an average between 300 and 500\,kms\sur{-1}, depending on the solar cycle phase, and they carry enormous amounts of mass and energy, of the order $10^{12}$\,kg and $10^{25}$\,J, respectively \citep{chen11,webb_howard12}.  

CMEs are often, but not always, associated with flares, namely abrupt, localized brightenings of the solar atmosphere, observed throughout the entire electromagnetic spectrum \citep{fletcher11,shibata11}. The flare--CME association rate increases with peak emission and duration of the flare \citep{yashiro06,yashiro09}. The intensity of flares is categorized by their peak soft X-ray output at 1--8\,{\AA}, in a logarithmic scale of classes (X, M, C, B, and A, in decreasing intensity), complemented by decimal subclasses. Flares also affect space weather either through $\gamma$-ray, X-ray, and EUV emission or particle radiation. Particles can be accelerated both at flaring sites and at the fronts of propagating CMEs, resulting in solar energetic particle (SEP) events \citep{dierckxsens15,papaioannou16}. Disentangling and mitigating the effects of these (chains of) events in our space-dependent modern civilization requires close monitoring and detailed understanding of the CME initiation process. 

CMEs and associated eruptive flares are manifestations of energetic phenomena powered by the magnetic energy stored in solar magnetic configurations. When magnetic flux emerges to form active regions, manifested as sunspot groups in white light \citep{vandrielgesztelyi09}, magnetic energy accumulates in the solar atmosphere. The amount of active-region magnetic energy due to electric currents (\textit{i.e.} their non-potentiality) is evident in coronal and chromospheric observations by the twist of magnetic structures pervading them \citep{leka96,schrijver16}. Part of this stored energy is amply capable of driving CMEs and flares.

The most important morphological feature of active regions that contain large amounts of free energy, and are thus prone to erupt, is the presence of one or more strong magnetic polarity-inversion lines (MPIL). Intense MPILs mark regions of opposite magnetic polarity in very close proximity and are generally associated with strong shearing motions \citep[see, \textit{e.g.},][and references therein]{park18}.  The strength or ``importance'' of a MPIL is usually parameterized through quantities calculated from photospheric magnetograms. These can be flux-weighted measures of MPIL length \citep[\textit{e.g.} $WL$\sous{SG}: ][]{falconer08}, MPIL-associated magnetic flux \citep[\textit{e.g.} $R$: ][]{schrijver07}, net electric currents producing or being produced by shear along MPILs \citep{georgoulis12a,kontogiannis17}, or other morphological properties assuming a certain connectivity between opposite polarities \citep[such as $B$\sous{eff} or Ising energy; ][]{georgoulis07,ahmed10}. Although associated with MPILs, these parameters contain distinct, independent information \citep{kontogiannis18}. These quantities, along with many others, are now being routinely used in flare and CME prediction schemes \citep{Leka_barnes07,Bobra_couvidat,florios18} with relative efficiency tested on larger and more detailed data sets. 

The impact of CMEs on space weather depends on their propagation speed and magnetic structure (\textit{e.g.} orientation of the magnetic field). In a more general sense, these quantities apply to the arrival of a CME at any point in the heliosphere (\textit{i.e.} natural bodies or spacecraft). The magnetic structure of a CME determines the specific interaction between the CME and the planetary or spacecraft magnetospheres or atmospheres.
Predicting or even studying the evolution of the CME as it propagates through interplanetary space can be a challenging task that requires a combination of \textit{in-situ} and remote-sensing measurements with theoretical modelling \citep[see, \textit{e.g.}, ][]{vourlidas13,palmerio17}. In contrast, research regarding the prediction of CME arrival times and impact speeds has advanced significantly during the past few decades with a number of models that describe and project the propagation of CMEs in the interplanetary space \citep[\textit{e.g. }][]{zhao_dryer14,mostl14,mays15}. These models vary in sophistication, but most of them share the same main inputs, in particular the initial kinematic characteristics of CMEs inferred by coronagraphic observations. 

Therefore, important goals of space-weather forecasting would be to, first, pinpoint those active-region characteristics best suited for CME prediction and, second, link them to the kinematic characteristics of CMEs. Intensive (\textit{i.e.} related more to active-region complexity than size) physical parameters have been found to be better suited for CME prediction \citep{bobra16}. Recent simulations appear to corroborate these results: for example, \citet{guennou17} show that naturally intensive parameters related to MPILs are the most promising predictors of CMEs. Earlier studies have also shown that there may be a correlation between the CME speed and some measure of magnetic energy \citep{venkatakrishnan03}, reconnection flux \citep{qiu05}, or effective connected magnetic-field strength \citep{georgoulis08}. Correlations with CMEs improve when parameters related to the MPIL alone, not the entire active region, are taken into account  \citep{vasantharaju18,pal18}.

In the era of regular high-quality solar and heliospheric observations, detailed catalogues and databases, combining information from all available sources may serve the purpose of CME prediction. This was showcased by \cite{murray18} who combined CME detections and characteristics from the EU Framework Package 7 HELCATS project (\url{www.helcats-fp7.eu}) with active-region magnetic properties from the EU Horizon 2020 FLARECAST (\url{flarecast.eu/}) project. Although correlations tend to weaken in larger samples, general trends persist and, moreover, an upper limit on CME speeds may be imposed by magnetic non-potentiality parameters \citep{tiwari15}.  

Motivated by these works, we compare several MPIL-associated quantities on grounds of their association with kinematic characteristics of CMEs. Some of these parameters are well-established, already in use in flare forecasting, while others have been developed/implemented recently for the FLARECAST project, and their association with CMEs is tested here for the first time. The comparison applies to a sample for which the active-region CME sources are unambiguously determined. Particular attention is given to fast CMEs, as these are mostly important in terms of geoeffectiveness \citep{sheeley99}.

\section{Data and Sample Selection}
\label{S-data}
For this study we selected a sample of flare-associated CMEs whose source regions are unambiguously identified, combining information and data from three online, widely used databases.

The Space Weather Database of Notifications, Knowledge, Information (DONKI: \url{kauai.ccmc.gsfc.nasa.gov/DONKI/}) is a tool provided by the Community Coordinated Modeling Center (CCMC), developed at the Goddard Space Flight Center (GSFC) of the National Aeronautics and Space Administration (NASA). The DONKI database contains a catalogue of space-weather phenomena such as flares, CMEs, and SEPs, along with all available information from simulations, modelling, and prediction. We used the DONKI flare catalogue to associate flares and CMEs with known National Oceanic and Atmospheric Administration (NOAA) Active Regions (AR), which are also included (if determined) into the catalogue. 

The \textit{Large Angle and Spectrometric Coronograph} (LASCO) Coordinated Data Analysis Workshop (CDAW) catalogue \citep{gopalswamy09} was used to extract the kinematic characteristics of CMEs. Height--time measurements are fitted to produce the (plane-of-sky) CME speeds and accelerations, while white-light observations combined with several assumptions can lead to an estimation of the CME (representative) mass and kinetic energy. Observational constraints on white-light detections of CME fronts may not always allow the determination of acceleration, kinetic energy, and/or mass \citep{vourlidas02}. Events missing one of these inferences were discarded. Since these constraints do not depend on the speed but on the visibility of the leading front of the CMEs \citep[see e.g.][]{yashiro04}, discarding these events should not impose any bias on the sample.

The properties related to MPILs were calculated from vector magnetograms taken by the \textit{Helioseismic Magnetic Imager} \citep[HMI:][]{hmischerrer,hmischou} onboard the \textit{Solar Dynamics Observatory} mission \citep[SDO:][]{sdo}. The Space weather HMI Active Region Patches \citep[SHARP:][]{bobra14} were developed specifically for use in space-weather forecasting applications. These cut-outs contain, among others, maps of the magnetic-field vector components of areas of interest, $B_{r}$, $B_{p}$ and $B_{t}$, and their corresponding errors, $B_{r,}$\sous{err}, $B_{p,}$\sous{err} and $B_{t,}$\sous{err}. These are deprojected and remapped into cylindrical equal area (CEA) map coordinates, as if observed at the solar disk center, with each pixel of the map corresponding to a fixed area at the photosphere \citep[see ][and references therein for further details]{bobra14}. 

SHARPs are usually, but not always, accompanied by NOAA AR identifiers but the correspondence is not one-to-one since a cut-out may contain one, several, or no NOAA-numbered active regions. This is the case for active regions that are too close to each other to be assigned to different cut-outs by the automatic segmentation algorithm. In such cases, the MPIL-related parameters that are calculated for the entire cut-out do not characterize the actual source region of the event. To avoid erroneous correspondences between source region and MPIL-parameter values, the respective events and source regions were rejected (e.g. the M7.6 eruptive flare from NOAA AR 12565, since HARP 6670 contained both NOAA AR 12565 and 12567). Furthermore, we only study eruptions for which there exist observations for the entire day (24 hours) preceding them. The deprojected, remapped vector magnetograms provide an opportunity to calculate active-region parameters with no restrictions on disk position. However, there are dependencies of the MPIL-related parameters on the position on the solar disk due to instrumental (\textit{e.g.} noise level of the HMI data) and projection effects (\textit{e.g.} foreshortening of active regions and/or false magnetic polarities). The impact of these effects depends on the particular way each property relies on the measurements of the magnetic-field components \citep{guerra18,kontogiannis18}. Although we impose no restrictions on disk position, for some active regions located too close to the solar limb it was not possible to calculate MPIL-related properties. The corresponding events were also discarded from our sample.

After applying the above selection criteria, we ended up with 32 eruptive flares (Table~\ref{Table:t1}) stemming from 22 unambiguously identified source regions, spanning the SDO era in Solar Cycle 24. These are divided into fast and slow events using a threshold equal to 750\,kms\sur{-1} \citep{sheeley99}. A short discussion on the selection of this speed threshold is provided in the Appendix.

\begin{table}
\caption{Our eruptive flare sample and supporting information (see Section.~\ref{S-data})}
\setlength{\tabcolsep}{4.3pt}
\begin{tabular}{lccccc}
\hline
Flare & NOAA& GOES & Heliographic & CME  & CME \\
peak time  &  & Flare class & location & Linear Speed & Width \\
  &  &             &          &    [kms\sur{-1}]    &  [deg] \\ 
\hline
2011-02-15T01:44:00.000 & 11158 & X2.2 & S20W10 &  669 &    360 \\ 
2012-03-05T03:30:00.000 & 11429 & X1.1 & N19W58 &  1531 &    360 \\
2012-03-07T00:02:00.000 & 11429 & X5.4 & N18E31 &  2684 &    360 \\
2012-03-07T01:05:00.000 & 11429 & X1.3 & N17E27 &  1825 &    360 \\
2013-04-11T06:55:00.000 & 11719 & M6.5 & N07E13 &  861 &    360 \\
2013-10-22T21:15:00.000 & 11875 & M4.3 & N05E03 &  459 &    360 \\
2013-10-28T01:41:00.000 & 11875 & X1.0 & N07W63 &  695 &    360 \\
2013-10-28T04:32:00.000 & 11875 & M5.1 & N07W65 &  1201 &    156 \\
2013-11-08T04:20:00.000 & 11890 & X1.1 & S10E11 &  497 &    360 \\
2013-11-10T05:08:00.000 & 11890 & X1.1 & S11W16 &  682 &    262 \\
2014-01-07T18:02:00.000 & 11944 & X1.2 & S15W10 &  1830 &    360 \\
2014-02-11T03:22:00.000 & 11974 & M1.7 & S13E16 &  222 &    81 \\
2014-03-29T17:36:00.000 & 12017 & X1.0 & N10W32 &  528 &    360 \\
2014-04-25T00:17:00.000 & 12035 & X1.3 & S14W29 &  456 &    296 \\
2014-10-24T07:37:00.000 & 12192 & M4.0 & S19W05 &  677 &    96 \\
2014-11-07T16:53:00.000 & 12205 & X1.6 & N15E35 &  211 &    67 \\
2015-03-09T23:29:00.000 & 12297 & M5.8 & S17E39 &  995 &    360 \\
2015-03-10T03:19:00.000 & 12297 & M5.1 & S16E38 &  1040 &    360 \\
2015-03-11T16:11:00.000 & 12297 & X2.2 & S16E26 &  75 &    110 \\
2015-03-15T01:15:00.000 & 12297 & C9.1 & S22W29 &  719 &    360 \\
2015-06-18T16:33:00.000 & 12371 & M3.0 & N15E45 &  1305 &    360 \\
2015-06-22T17:39:00.000 & 12371 & M6.5 & N13W05 &  1209 &    360 \\
2015-08-21T09:34:00.000 & 12403 & M1.4 & S17E26 &  555 &    270 \\
2015-08-22T06:39:00.000 & 12403 & M1.2 & S15E13 &  547 &    360 \\
2015-09-20T17:32:00.000 & 12415 & M2.1 & S22W50 &  1239 &    360 \\
2015-10-22T02:13:00.000 & 12434 & C4.4 & S10W34 &  817 &    360 \\
2015-11-04T03:20:00.000 & 12445 & M1.9 & N14W65 &  272 &    64 \\
2015-11-04T13:30:00.000 & 12443 & M3.7 & N06W04 &  578 &    360 \\
2015-12-01T07:59:00.000 & 12458 & C3.6 & N10W42 &  333 &    71 \\
2015-12-16T08:34:00.000 & 12468 & C6.6 & S14W02 &  579 &    360 \\
2015-12-28T11:20:00.000 & 12473 & M1.8 & S22W12 &  1212 &    360 \\
2016-04-18T00:14:00.000 & 12529 & M6.7 & N10W51 &  1084 &    162 \\
\hline
\end{tabular}
\label{Table:t1}
\end{table}

\section{Parameters Associated with MPIL Strength}
For the 32 NOAA ARs of Table~\ref{Table:t1} we calculated ten properties associated with MPIL strength. These properties have been proposed in the literature as potential flare predictors and comprise a subset of the predictors studied in the course of the FLARECAST project. All properties were calculated from vector magnetograms (either the normal field component [$B_{r}$] or all three components of the magnetic field vector),  for the 24-hour window preceding the CME-associated flare with a cadence of 1 hour. Here, we give a brief description of each parameter, while a summary can be found in Table~\ref{Table:t2}. All algorithms for the calculation of the parameters listed in Table~\ref{Table:t2} were  created or tested/scrutinized by the FLARECAST consortium, based on the original works that introduced them. These algorithms are publicly available at the project's algorithm repository (\url{dev.flarecast.eu/stash/projects}).

\subsection{Effective Connected Magnetic Field Strength}
\label{S:beff}
The effective connected magnetic-field strength [$B\sous{eff}$], first introduced by \citet{georgoulis07}, is a morphological parameter that quantifies strong MPILs. Here we use the modified version of $B\sous{eff}$ described by \citet{georgoulis13}. The map of the radial component of the magnetic field, $B_{r}$, is partitioned into non-overlaping unipolar patches \citep{barnes05}, using thresholds of the magnetic-field strength ($B\sous{thres}=200$\,G), magnetic flux content per partition ($\Phi\sous{thres}=5 \times 10^{19}$\,Mx) and partition area (40\,pixel). Then, by means of a simulated annealing process, a connectivity matrix is calculated, with elements representing the magnetic flux assigned to the connection between any given pair of opposite-polarity partitions. With the connectivity matrix determined, $B\sous{eff}$ is the sum of all magnetic-flux elements, weighted by the squared length of each connection (\textit{i.e.} the distance between pairs of opposite polarity partitions) $L_{ij}$: 

\begin{equation}
B\sous{eff}=\sum_{i}\sum_{j}\frac{\Phi_{ij}}{L^{2}_{ij}}.
\label{equation:beff}
\end{equation}

\noindent The two sums shown in Equation~\ref{equation:beff} span over the number of positive- and negative-polarity partitions, respectively.

\subsection{Non-Neutralized Electric Currents}
\label{S:nn_cur}
The non-neutralized electric currents quantify the amount of net current injected into the corona. Strong non-neutralized currents are exclusively linked to the presence of strong MPILs \citep{georgoulis12a,torok14,dalmasse15}. We calculate $I\sous{NN,tot}$ and $I\sous{NN,max}$ according to the process described by \citet{kontogiannis17}. The map of the radial component of the SHARP vector magnetic field is partitioned into magnetic patches (see Section~\ref{S:beff}). For each partition, the vertical electric-current density is calculated \textit{via} Amp\`{e}re's law,

\begin{equation}
J_{z}=\frac{1}{\mu_{0}}(\frac{\partial B_{y}}{\partial x}-\frac{\partial B_{x}}{\partial y}),
\label{equation:ampere}
\end{equation}

\noindent where $B_{x}$ and $B_{y}$ are the horizontal components of the magnetic field and $\mu_{0}$ is the magnetic permeability of vacuum. The algebraic sum of $J_{z}$ contained within each partition corresponds to the total electric current of this partition. The uncertainty of this total electric current per partition is calculated \textit{via} error propagation, using the error maps of the two horizontal components of the magnetic field. Then, the radial component, $B_{r}$ is used to calculate the components of the corresponding potential magnetic field at the photosphere \citep{alissandrakis81}. With these we calculate again the total current within each partition. Although by definition this current should be equal to zero in case of potential fields, numerical effects may result in non-zero total currents. A magnetic partition is considered non-neutralized only when its total electric current exceeds its uncertainty and the numerical current of the potential magnetic field by factors of three and five, respectively. For an ensemble of non-neutralized partitions, each containing a total current $I_{i}\sur{NN}( i \equiv {1,...n})$, we calculate the maximum unsigned value $I\sous{NN,max}=max{|I_{i}\sur{NN}|}$ and the total unsigned non-neutralized current $I\sous{NN,tot}= \sum_i {|I_{i}\sur{NN}|}$. Therefore, we can calculate the total non-neutralized currents per magnetic polarity in any given active region, which is a quantity found to depend on MPILs \citep{georgoulis12a}.

\subsection{Ising Energy}
\label{S:ising}
\noindent The Ising energy is a measure of the compactness of an active region, introduced by \citet{ahmed10} as a promising flare predictor. The Ising energy of an ensemble of opposite magnetic polarity pixels is given by

\begin{equation}
E\sous{Ising}=-\sum_{i,j}\frac{S_{i}S_{j}}{d^{2}},
\label{equation:ising}
\end{equation}

\noindent where $S_{i}$ and $S_{j}$ denote the polarity of the magnetic field at pixels \textit{i} and \textit{j} (+1 for positive and -1 for negative), and \textit{d} is the separation distance between these two pixels. Only strong magnetic-field pixels (with field $> 200\,G$) are taken into account. \citet{kontogiannis18} tested the efficiency of this predictor in a representative sample of SHARP data and found that it could be useful in automated flare-prediction services. They also proposed two variations, namely the Ising energy of the magnetic partitions (which are already determined for $B\sous{eff}$) and the Ising energy of the sunspot umbrae (which also requires continuum observations). The former, $E\sous{Ising,part}$,  is further utilized in this study.

\subsection{Schrijver's $R$}

Schrijver's $R$ quantifies the unsigned magnetic flux near strong, high-gradient MPILs \citep{schrijver07}. It is calculated from the $B_{r}$-map and is based on a MPIL detection algorithm. Masks of strong positive- and negative-polarity magnetic field are determined ($|B_{r}| > 300$\,G). Then, these masks are dilated by a 3$\times$3 kernel. The areas where positive and negative masks overlap define the strong MPILs. The subsequent mask that contains the MPILs is convolved with a $\approx$15\,Mm-wide 2D Gaussian kernel and then multiplied with the map of the radial magnetic-field component. The sum of the absolute values of this composite map is the $R$-value, whose base-ten logarithm is used for flare prediction.

\begin{table}
\caption{Parameters characterising MPILs in this study.}
\setlength{\tabcolsep}{4.3pt}
\begin{tabular}{ m{2.75cm} m{2.75cm} m{2.75cm} m{2.75cm} }
\hline
Parameter& Represents& Input& Reference\\
\hline
$I\sous{NN,tot}$/$I\sous{NN,max}$  & Net currents injected in the corona & Vector        magnetogram & \citet{kontogiannis17} \\ 
$B\sous{eff}$ & Magnetic connectivity, favouring MPILs & $B_{r}$ & \citet{georgoulis07} \\
Ising Energy & Compactness & $B_{r}$ & \citet{ahmed10} \\
R value & MPIL magnetic flux& $B_{r}$ & \citet{schrijver07} \\
$WL\sous{SG}$ & \begin{flushleft}weighted MPIL length\end{flushleft}& $B_{r}$ & \citet{falconer08}\\
\begin{flushleft}MPIL length, total/maximum\end{flushleft} & MPIL length& $B_{r}$ & \citet{mason_hoeksema10}\\
MPIL flux & MPIL magnetic flux& $B_{r}$ & \citet{mason_hoeksema10}\\
\hline
\end{tabular}
\label{Table:t2}
\end{table}

\subsection{Gradient-Weighted Integral Length of Neutral Line [$WL\sous{SG}$]}

Another way to parameterize the strength of the MPIL in an active region is its gradient-weighted integral length introduced by \citet{falconer08}, in use in the MAG4 flare prediction service (\url{www.uah.edu/cspar/research/mag4-page}). This parameter is the integral along the neutral line of the horizontal gradient of the vertical component of the magnetic field: 

\begin{equation}
WL\sous{SG}=\int (\nabla B_{z})\textrm{dl}.
\label{equation:wlsg}
\end{equation}

\noindent The detection of MPILs is based on the process described by \citet{falconer03}: images of the vertical magnetic-field component are heavily smoothed and possible MPILs are identified as contours of zero magnetic field. Then the horizontal components of a potential magnetic field are calculated from these images \citep{alissandrakis81}. Thresholds on the gradient of the vertical magnetic field and the strength of the horizontal magnetic field are used to determine the MPILs. The process is repeated for less-smoothed input images and comparison of the results leads to the determination of the MPILs. A detailed description of the process and its application to SHARP data has been given by \citet{guerra18}.  
  
\subsection{Magnetic Polarity Inversion Line Length and Associated Magnetic Flux}

An MPIL detection method similar to that employed for the calculation of the $WL\sous{SG}$ can also be used to determine more rudimentary characteristics of MPILs \citep{mason_hoeksema10}. These are the total length of all detected MPILs, $MPIL\sous{tot}$ the length of the longest (main) MPIL, $MPIL\sous{max}$, and the total unsigned magnetic flux near MPILs, $MPIL\sous{flux}$. It should be noted that $MPIL\sous{flux}$ differs from $R$ both in the detection method of MPIL and the distance over which it is calculated (2\,Mm instead of 15\,Mm). The MPIL related parameters have also been included in the list of predictors utilized by the FLARECAST service. Application of these predictors to a sizeable sample of SHARP data has been described by \citet{guerra18}.

\subsection{Parameter Uncertainties}
\label{Section:uncert}

Assessing the uncertainty of each parameter is important to judge whether any observed temporal variations of the parameters are meaningful. For the non-neutralized currents, this calculation incorporates the measurement uncertainties of the three components of the magnetic field. For other properties, such as $B\sous{eff}$, the Ising energies and the MPIL length, producing analytical expressions for the uncertainties is not possible.

To provide an estimate of the uncertainties propagated to the MPIL-related values, we use the following methodology:
in a single map of the radial component of the magnetic field of NOAA AR\,11158 we produce 1000 random realizations, assuming that each pixel has a value equal to its original plus a random value that follows a normal distribution with a standard deviation equal to 100\,G. This value was selected because 99.9\,\% of the values in the error map are found below 100\,G. We then calculate the MPIL-related properties of the 1000 random realizations and fit a Gaussian to each distribution of property values. The standard deviation of this Gaussian is considered an estimated uncertainty.
This process was not performed for $I\sous{NN,tot}$ and $I\sous{NN,max}$, whose uncertainties were calculated via normal error propagation. In principle, uncertainties calculated for smaller active regions may be higher than those of larger active regions. However, since our sample contains only eruptive active regions, which are large and well-developed, the uncertainties calculated here are representative of our sample.

 \begin{figure}
 \centerline{\includegraphics[width=1\textwidth]{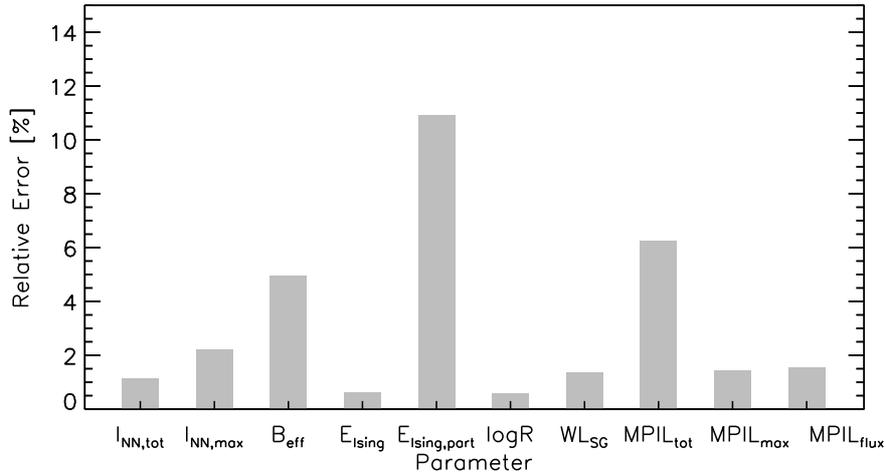}}
 \caption{Calculated ($I\sous{NN,tot}$ and $I\sous{NN,max}$) and estimated (all remaining properties) relative uncertainties for the ten MPIL-related parameters of this study.}
\label{fig:errors}
 \end{figure}

The results of this process are summarized in Figure~\ref{fig:errors}. All relative uncertainties are well below 10\,\%, except for the Ising energy of the partitioned magnetograms, $E_{Ising,part}$ (11\,\%). For this parameter, fluctuations of the radial magnetic-field component may affect the number of partitions produced by the tessellation scheme, and since the number of partitions within an active region is usually of the order of tens or hundreds, the value of the parameter can be influenced more than other parameters of the study. In contrast, $E\sous{Ising}$ comprises the sum of all pixels above a certain threshold and, therefore, it is less sensitive to the fluctuations of the radial component. Similarly, the effect of these fluctuations on $B\sous{eff}$, which also relies on $B_{r}$-partitions, is significantly alleviated because the connections and the contained magnetic fluxes of the partitions are also taken into account for the calculation. 

$MPIL\sous{flux}$, $MPIL\sous{max}$, $R$ and $WL\sous{SG}$ represent the sum of magnetic flux or gradient in a large number of pixels around the MPIL and therefore the fluctuations of the radial magnetic field within each pixel are smoothed out. These fluctuations can, in principle, affect the determination of the MPIL region, but they appear to affect less the length of the main neutral line. Even more so in the case of $R$ which is, in essence, the base-ten logarithm of the MPIL region magnetic flux.

Finally, regarding the non-neutralized currents ($I\sous{NN,tot}$ and $I\sous{NN,max}$) their measurement is based already on the condition that in each partition non-neutralized currents should exceed three times the corresponding uncertainty and five times the numerical current while their calculation is  based on a large number of pixels. For these reasons, the relative errors for $I\sous{NN,tot}$ and $I\sous{NN,max}$ are generally low \citep[see also the appendix of][]{kontogiannis17}.

 \begin{figure}
 \centerline{\includegraphics[width=1\textwidth]{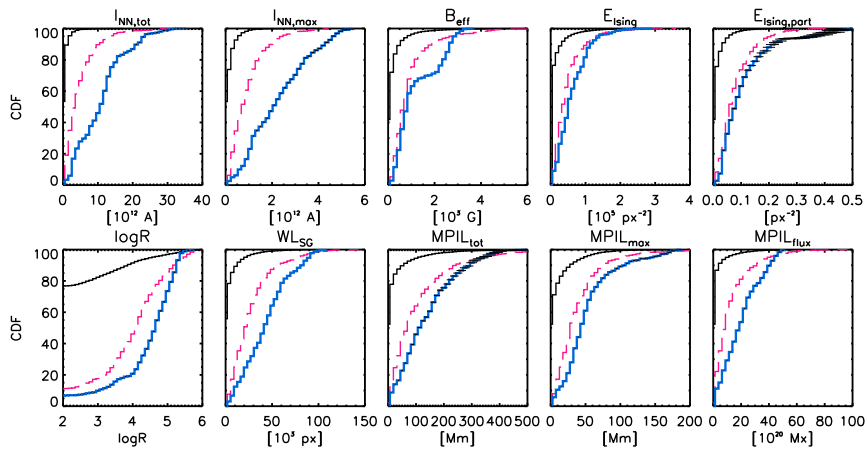}}
 \caption{Cumulative density functions (CDF) of MPIL-related properties for the three samples of active regions discussed in Section~\ref{Section:control}: CDFs of parameters stemming from the non-flaring part of the random sample are shown in \textit{solid black}. CDFs of parameters stemming from the flaring (above C1.0) part of the random sample are shown in \textit{dashed purple}, while CDFs stemming from the 22 active regions of this study are shown in \textit{thick blue}. For the latter, representative uncertainties, as per Figure~\ref{fig:errors}, are indicated as \textit{black horizontal bars}. These errors are smaller than the thickness of the lines, except for $E\sous{Ising,part}$ and $MPIL\sous{tot}$. }
\label{fig:cdf}
 \end{figure}

\section{Results}
 
\subsection{Parameter Values of the Eruptive Active Regions}
\label{Section:control}

A parameter useful for flare and CME prediction should allow, in a probabilistic way, the separation between flaring and non-flaring, or between eruptive and non-eruptive active regions. For flares, this has been already demonstrated for various data sets \citep[see the references included in the last column of Table~\ref{Table:t2} and][]{guerra18,kontogiannis18}.

For consistency, here we compare the values of the MPIL-related properties during the 24-hour window that preceded the 32 CMEs in this study with those of a random sample of 9454 point-in-time SHARP cut-outs. We randomly selected 25\,\% of the days between September 2012 and May 2016 and for each day we used all SHARPs with a six-hour cadence. This sample, which was also used by \citet{kontogiannis18} consists of 1021 flaring ($>$C1.0) and 8433 non-flaring regions. For flaring regions, flares occurred within 24 hours from the time that the parameter values were taken.

In Figure~\ref{fig:cdf} we plot the cumulative distribution functions (CDF) for the three samples (flaring, non-flaring, and the one of the present study). All CDFs were calculated similarly, by considering 40 threshold values, uniformly distributed along their dynamic range. For all parameters, there is a clear segregation between flaring (red dashed) and non-flaring regions (black solid), the former being shifted towards higher values, as expected. The parameter values for the eruptive active regions studied here (blue thick) are further shifted towards the high end of the distributions, which means that, statistically, higher values of these parameters are observed for active regions that produce CMEs within the following day.

We note that CME prediction is outside the scope of this study. Addressing this subject would require an extended dataset and sophisticated statistical methods. However, from Figure~\ref{fig:cdf} we conclude that the CME-productive active regions of our sample exhibit clearly and consistently higher (beyond the error margin) MPIL-related values compared to quiet and non-eruptive flaring active regions. The minimum separation between eruptive and non-eruptive active regions is found for the values of the Ising energies while the maximum is found for the two non-neutralized electric-current parameters. The statistical separation of property values for different populations of active regions is an important research topic and we will further investigate this in subsequent studies. 

\begin{figure}
\centerline{\includegraphics[width=0.6\textwidth]{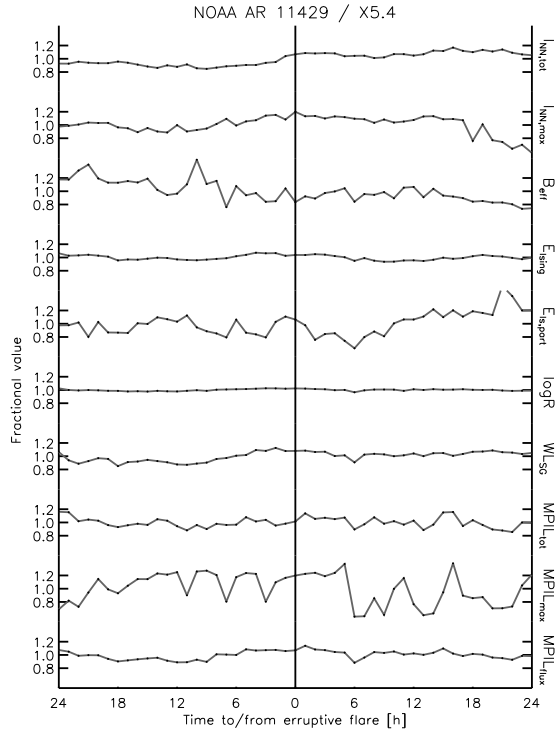}}
\caption{Temporal variation of the ten MPIL-related parameters during the 24 hours that preceded the X5.4 flare produced by AR\,11429. All values are given as fraction of the corresponding (temporal) average value.}
\label{fig:time_series}
\end{figure}

\begin{figure}
\centerline{\includegraphics[width=1\textwidth]{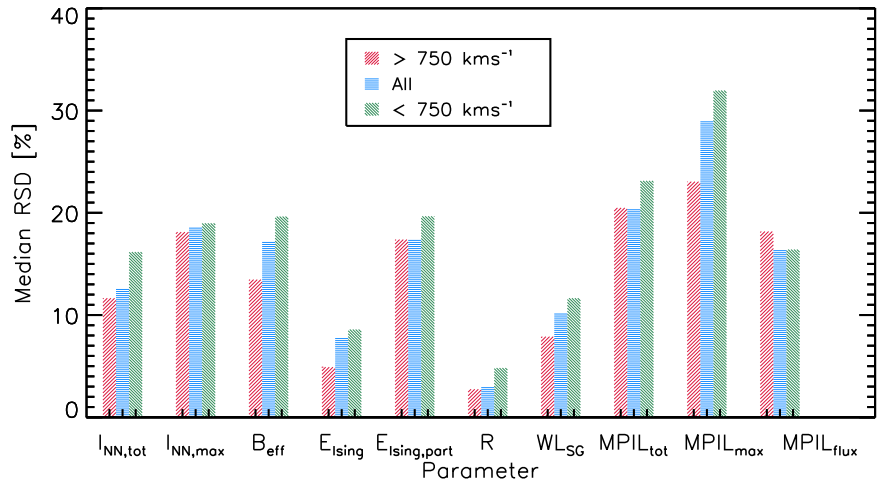}}
\caption{Median of the Relative Standard Deviation (RSD) of each parameter time series.}
\label{fig:rsd}
 \end{figure}

\subsection{Parameter Temporal Variability and Correlations}
\label{Section:temp}

In Figure~\ref{fig:time_series} we plot the evolution of the ten MPIL-related properties of NOAA AR\,11429, during the 24 hours that preceded the X5.4 flare on 7 March, 2012 and during the 24 hours that followed. This was one of the strongest flares of Solar Cycle 24, accompanied by a very fast CME. Both the source region and corresponding events have been studied extensively \citep[see, e.g.,][]{patsourakos12,syntelis16}. 

Although all properties are associated with the non-potentiality of active regions, their temporal evolution towards and after the eruption is far from identical. Most parameters range within 20\,\% of their daily averages. Overall, the highest variability in Figure~\ref{fig:time_series} is exhibited by the $E\sous{Ising,part}$ and the $MPIL\sous{max}$. As indicated also by their uncertainties in Figure~\ref{fig:errors}, these parameters are more sensitive to radial magnetic-field variations from map to map. On the other hand, the value of $logR$ changes barely: only within the error margins. Some parameters continue to increase for several hours after the eruption before they start decreasing ($I\sous{NN,tot}$, $MPIL\sous{max}$). Others, such as $B\sous{eff}$ and $E\sous{Ising,part}$, exhibit distinct, beyond the error margin, peaks before the eruption and a larger variation. Regarding $B\sous{eff}$, a similar behaviour was also found in NOAA AR\,11158 by \citet{georgoulis13}, where the peak occurred a few hours before the X-class flare. As also suggested by Figure~\ref{fig:cdf}, all parameters have consistently high values during the 24 hours that precede eruptive flares. However, the outlined differences between their temporal evolution result in weak correlations between most of the parameters, during the studied interval, suggesting that these do not necessarily contain redundant information.

The evolution presented in Figure~\ref{fig:time_series} is a representative example, although there can be differences from active region to active region, in the percentages within which each parameter ranges, and in the position of the peaks, indicating a far-from-unique behaviour of different MPIL-related properties as active regions evolve towards eruptions. Also, these parameters have different dependencies on instrumental uncertainties because of their different dependence on magnetic-field measurements \citep[see Section~\ref{Section:uncert} and also][]{kontogiannis18}. 

Extending this preliminary analysis on the temporal variability of the MPIL-related properties prior to major events over the entire eruptive flares sample, we calculate the relative standard deviation (RSD) of every parameter for each time series during the 24 hours preceding each event. For the sample of the eruptive active regions, we plot the median RSD of each parameter in Figure~\ref{fig:rsd}. Overall, the median RSD is below 30\,\%, with the maximum of $\approx$30\,\% being exhibited by $MPIL\sous{max}$ and the minimum, $\approx$2\,\% by $logR$. An interesting finding is that time series associated with fast events exhibit almost invariably a lower median RSD, meaning that the temporal evolution of these MPIL-related parameters as the active region evolves towards a fast CME is relatively smaller, possibly because these parameters tend to have continuously larger values for stronger MPILs. In the 24 hours following eruptions (not shown), median RSD differ by $\approx$5\,\% but the relation between the median RSD of fast and slow events is preserved. This indicates that over the day following eruptions there are no dramatic changes in the values of the parameters, corroborating established knowledge that the line-tied photosphere changes on timescales significantly longer than the erupting, overlaying corona.

\begin{figure}
\centerline{\includegraphics[width=1\textwidth]{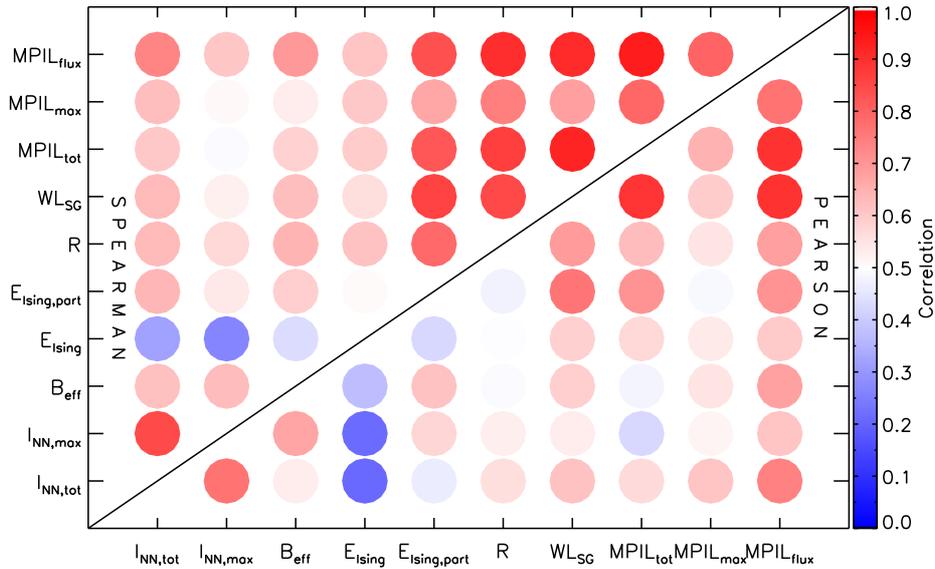}}
\caption{Correlations between the MPIL-related parameters as a bubble diagram, where the significance of each correlation is represented by the color of the bubble. The rank order (Spearman) and linear (Pearson) correlation coefficients are shown above and below the diagonal, respectively.}
\label{fig:param_correls}
\end{figure}
 
It has already been noted (discussing Figure~\ref{fig:time_series}) that for AR\,11429 correlations between MPIL-related properties are generally low during the 24 hours preceding the eruptive X5.4 flare. To complete the picture, we examine the correlations between the MPIL-related properties for the entire sample of active regions, by using each and every parameter value of the 32 time series. These correlations are presented in Figure~\ref{fig:param_correls} as a 10$\times$10 matrix of color-coded bubbles.
All parameters are positively correlated, reflecting the fact that for all properties higher values correspond statistically to higher flare probabilities. The positive correlation between quantities that are being used as predictors of flaring activity has been demonstrated by older \citep{Leka_barnes07} and more recent \citep{guerra18,kontogiannis18} studies. It has also been shown that most of these properties are positively correlated with the size of active regions, as expressed by the (extensive) total unsigned magnetic flux. The dependence on active-region size should be moderated for MPIL-related parameters as they are non-extensive but, since they are associated with the same topological feature, positive correlations are expected nonetheless. The highest correlation is found between similar parameters, \textit{e.g.} between $I\sous{NN,tot}$ and $I\sous{NN,max}$ (non-neutralized currents), between $R$, $MPIL\sous{flux}$ and $WL\sous{SG}$ (which are associated with magnetic flux at the vicinity of MPILs), etc. The parameter least correlated with the rest is $E\sous{Ising}$, but its ``partitioned'' variation exhibits higher correlation. The rank order (Spearman) correlation is overall slightly higher than the linear (Pearson) one. For example, we find moderate linear correlations between $R$ and $B\sous{eff}$--Ising energies but the Spearman correlation is stronger, implying that these properties are better correlated via a monotonic nonlinear relationship. For some cases (\textit{e.g.} between $E\sous{Ising}$ and $I\sous{NN,tot}$) correlations are lower than the ones presented by \citet{kontogiannis18} but it should be kept in mind that in the present study the MPIL-related properties are calculated from $B_{r}$ instead of $B_{LOS}$ (\textit{i.e.} the line-of-sight magnetic field) and, therefore, differences are anticipated \citep{guerra18}. Performing the same analysis for the daily average values of the parameter for each of the active regions, instead of using all points, leads to slightly stronger correlations between all parameters.

We hereafter focus on the average value of each parameter, calculated over the 24-hour window. Unlike the peak value, which may be subject to instrumental/numerical effects, we expect that the average will be a more reliable and robust measure, a common ground to compare different parameters with different types of evolution. Furthermore, although the present study does not address forecasting CMEs and their characteristics, we expect that a running-average value could be more practical in operational schemes than a peak value. 

\subsection{MPIL-Related Parameters and Flare Magnitude}

\begin{figure}
 \centerline{\includegraphics[width=1\textwidth]{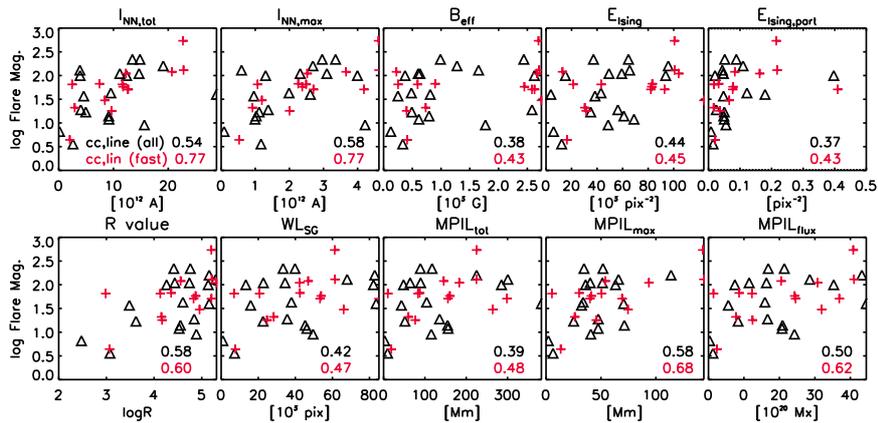}}
 \caption{Flare index \textit{vs} MPIL parameters. \textit{Black triangles} (\textit{red crosses}) refer to events where the CME linear speed was lower than (greater or equal to) 750\,kms\sur{-1}. The inset numbers are the corresponding Pearson correlation coefficients for all (\textit{top, black}) and fast events (\textit{bottom, red}).}
\label{fig:fmag_scat}
 \end{figure}

\begin{figure}
 \centerline{\includegraphics[width=1\textwidth]{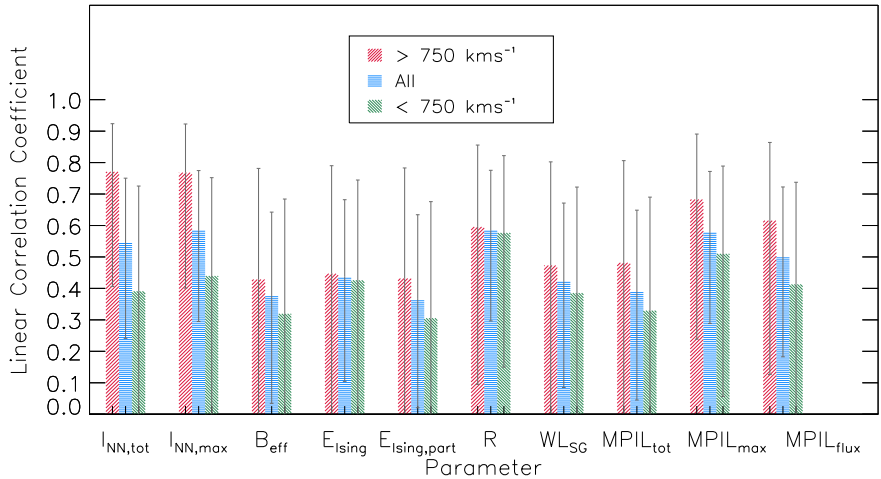}}
 \caption{Linear correlation coefficients between the average values of each parameter and the corresponding common logarithm of flare index. \textit{Error bars} mark the 95\,\% confidence intervals calculated \textit{via} a Fisher transformation.}
\label{fig:fmag_bar}
 \end{figure}

For each eruptive flare we calculate the corresponding flare index according to \citet{abramenko05}, although we do not sum over multiple flares, as was done in that work. In this representation the flare index of a X1.1 would be equal to 110, a M6.5 to 65, a C3.6 to 3.6, \textit{etc}. In Figure~\ref{fig:fmag_scat} we plot the flare index \textit{versus} the 24-hour-averaged property value prior to the event. Justifying their use as measures of active region non-potentiality, all properties exhibit positive correlations with the flare index. The higher the values of these parameters, the more prone to intense flaring are the source regions. In Figure~\ref{fig:fmag_bar} we plot the linear (Pearson) correlation coefficients calculated between the property 24\,h-average values and the flare magnitude. The Fisher transformation was used to calculate the 95\,\% confidence intervals and \citep[see e.g. ][]{press07}, which are indicated in the form of error bars. Regarding the entire dataset, the highest correlation is found for $logR$, followed by the non-neutralized electric currents, the main MPIL length [$MPIL\sous{max}$], and the associated flux [$MPIL\sous{flux}$] (Figure~\ref{fig:fmag_bar}, blue bars). The rest of the properties do not exhibit notable differences. However, the differences between all of the parameters are within the margins set by the 95\,\% confidence intervals. 

These differences are exacerbated when examining the correlations for fast CMEs (Figure~\ref{fig:fmag_bar}, red bars). The correlations increase for all parameters but the increase is larger for $I\sous{NN,tot}$, $I\sous{NN,max}$ and $MPIL\sous{max}$. On the other hand, for slow CMEs (Figure~\ref{fig:fmag_bar}, green bars) the correlations are overall lower. The Ising energy and the $R$-value are the predictors least affected by this sample selection, exhibiting more or less the same correlation with the flare index.

Another interesting point to consider is how these properties are grouped. The highest correlation is found for the current-related predictors, followed by the length of the main MPIL. Intermediate correlations are found for $R$ and $MPIL\sous{flux}$, which both quantify the amount of magnetic flux in the vicinity of strong MPILs, and $WL\sous{SG}$ and $MPIL\sous{tot}$ which refer to the length of all MPILs. Finally, the lowest correlations are for $B\sous{eff}$, $E\sous{Ising}$ and $E\sous{Ising,part}$, which involve also information on the connectivity of opposite-polarity pixels or partitions. As already mentioned, each of these parameters contains essentially different information: the relevance of different levels of information to flaring activity may be a root cause for the different correlations.

\subsection{MPIL-Related Parameters and CME Kinematic Characteristics}

The association between the 24-hour average values of the parameters and the linear speed of the CMEs is illustrated in Figure~\ref{fig:vspeed_scat}. These plots show basically what has been asserted in previous studies, namely that the maximum anticipated speed of a CME increases as the property value increases \citep{tiwari15}. For some properties (\textit{i.e.} $I\sous{NN,tot}$ and $MPIL\sous{max}$) investigated for the first time in this study, there is clearly a different behaviour for fast events compared to the slow ones. Speeds of faster events show a clear increasing trend with increasing MPIL-related property, contrary to slower events that are more or less distributed across the entire range of values implying a much weaker, if any, correlation. 

All properties are positively correlated with the CME speed, in line with results presented for a different set of parameters by \citet{murray18}, where a weak positive correlation was found between the polarity inversion line length, $R$, $WL\sous{SG}$, and the CME linear speed. Here, the highest correlations are found for $MPIL\sous{max}$, followed by non-neutralized currents for all CMEs.  

Correlations increase further for fast CMEs. In fact, it is the significantly lower correlation exhibited for the slow events (Figure~\ref{fig:cme_bar}a, green bars) that lowers the correlation coefficients of the entire sample. By means of a t-test \citep{press07}, we conclude that for $I\sous{NN,tot}$ and $MPIL\sous{max}$ the increase of the correlation coefficient for the fast events in comparison to the slow ones is statistically significant at 95\,\% level of significance. For fast CMEs, the highest correlation ($>$0.8) is exhibited by the total unsigned non-neutralized currents and the length of the longest neutral line $MPIL\sous{max}$. Also, these two parameters exhibit slightly stronger (but within errors) correlation with the CME speed than with the flare magnitude of fast events (see Figure~\ref{fig:fmag_bar}). One may obtain the following linear relations between $I\sous{NN,tot}$, $MPIL\sous{max}$ and the CME speed: 

 \begin{figure}
 \centerline{\includegraphics[width=1\textwidth]{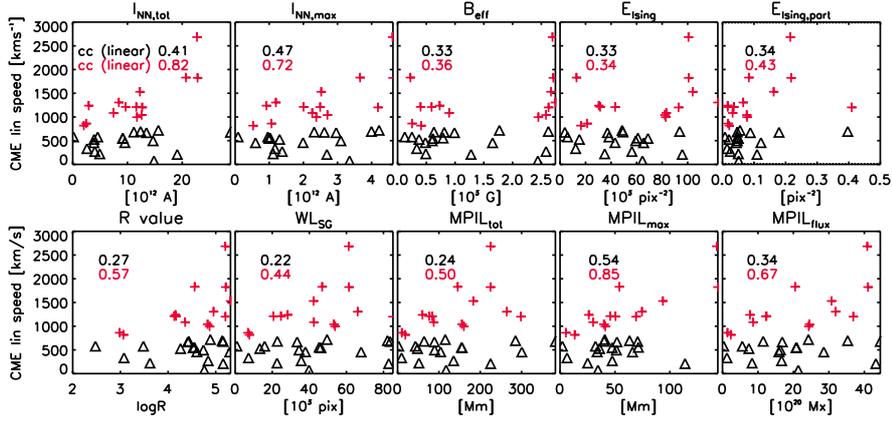}}
 \caption{CME linear speed \textit{vs} MPIL parameters. \textit{Black triangles} (\textit{red crosses}) refer to events where the CME linear speed was lower than (greater or equal to) 750\,kms\sur{-1}. The inset numbers are the corresponding Pearson correlation coefficients for all (\textit{top, black}) and fast events (\textit{bottom, red}).}
\label{fig:vspeed_scat}
 \end{figure}

 \begin{figure}
 \centerline{\includegraphics[width=1\textwidth]{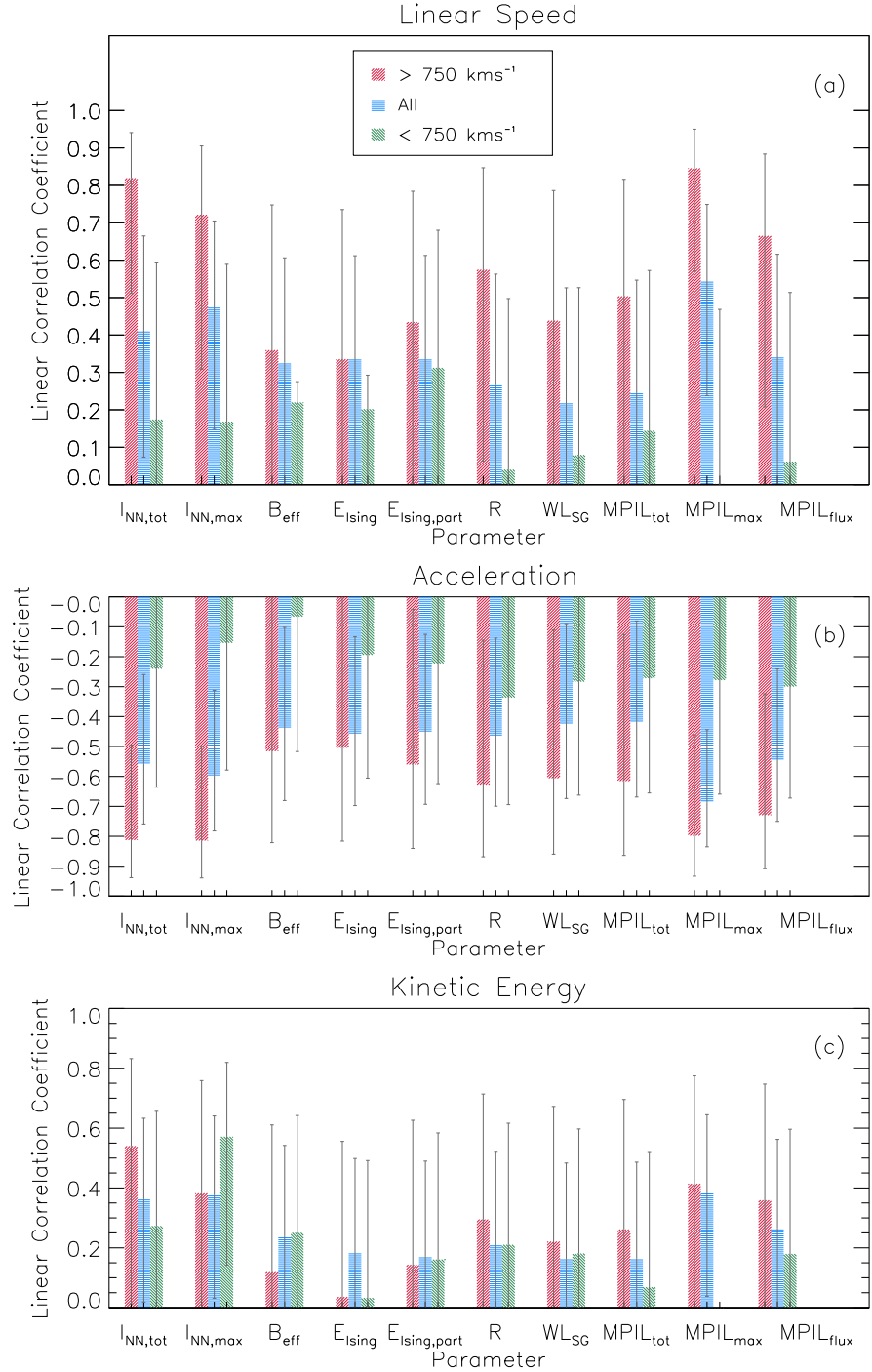}}
 \caption{Linear correlation coefficients between the 24-hour average values of each parameter and \textbf{(a)} the CME linear speed, \textbf{(b)} the CME acceleration, and \textbf{(c)} the CME kinetic energy. \textit{Error bars} mark the 95\,\% confidence intervals calculated \textit{via} a Fisher transformation.}
\label{fig:cme_bar}
 \end{figure}
  
%

\begin{equation}
V\sous{CME}=59(12)I\sous{NN,tot}+670(160),
\label{eq:v_vs_inn}
\end{equation}

\begin{equation}
V\sous{CME}=9.8(1.8)MPIL\sous{max}+760(130).
\label{eq:v_vs_mpil}
\end{equation}

\noindent where $V\sous{CME}$, $I\sous{NN,tot}$ and $MPIL\sous{max}$ are measured in kms\sur{-1}, $10^{12}$\,A and Mm, respectively, and the numbers in the parentheses denote the corresponding uncertainties. $I\sous{NN,tot}$ and $MPIL\sous{max}$ exhibit also a strong correlation with each other (0.58, see Figure~\ref{fig:param_correls}) and the slightly higher rank order correlation coefficient (0.62) points to a non-linear relationship. This means that they might be used interchangeably but, alternatively, one may use the following empirical bilinear form,


\begin{equation}
V\sous{CME}=27(19)I\sous{NN,tot} + 6(3)MPIL\sous{max}+670.
\label{eq:v_vs_inn_mpil}
\end{equation}

\noindent In fact, the combined correlation coefficient of CME linear speed and the linear combination of $I\sous{NN,tot}$ and $MPIL\sous{max}$ via Equation~\ref{eq:v_vs_inn_mpil} is slightly higher (0.87) than the individual ones (0.82 and 0.84). Of course, it is not recommended to use these \textit{a posteriori} derived relationships to predict the speed of imminent CME's since the prediction of flares and CMEs is probabilistic \citep[see, \textit{e.g.},][]{papaioannou15,anastasiadis17} and therefore a high value of these properties does not guarantee the initiation of a CME. In case there is a fast CME from a source region with certain values of $I\sous{NN,tot}$ and/or $MPIL\sous{max}$, however, on an independently achieved probability of CME occurrence, then Equations~\ref{eq:v_vs_inn}--\ref{eq:v_vs_mpil} could be used to project the CME's linear speed. 

Besides $I\sous{NN,tot}$ and $MPIL\sous{max}$, the magnetic flux associated with all neutral lines, $MPIL\sous{flux}$, also exhibits a strong correlation with the CME linear speed, followed closely by $logR$, whose correlation coefficient exceeds 0.55. 
   
In Figures~\ref{fig:cme_bar}b and \ref{fig:cme_bar}c we show the corresponding correlation coefficients for the other two kinematic characteristics of CMEs, namely the acceleration and the kinetic energy, respectively. It should be noted that the coefficients for acceleration are negative due to the known anticorrelation between CME speed and acceleration \citep[see, \textit{e.g.},][]{paouris17}. The general finding is again that the non-neutralized currents and the length of the main MPIL are the parameters best correlated with CME acceleration and kinetic energy, at least for fast CMEs. Different correlations for fast and slow CMEs also appear here but the difference between fast and slow CMEs are statistically significant only for $I\sous{NN,tot}$, $I\sous{NN,max}$ and $MPIL\sous{max}$. Correlation coefficients are generally lower for CME kinetic energy because this requires the CME mass, as well, which adds further uncertainty. For all parameters, the differences between fast and slow CMEs are not statistically significant. It is important to report that no appreciable correlation was found between any of the parameters and the CME mass.

 \begin{figure}
 \centerline{\includegraphics[width=0.6\textwidth]{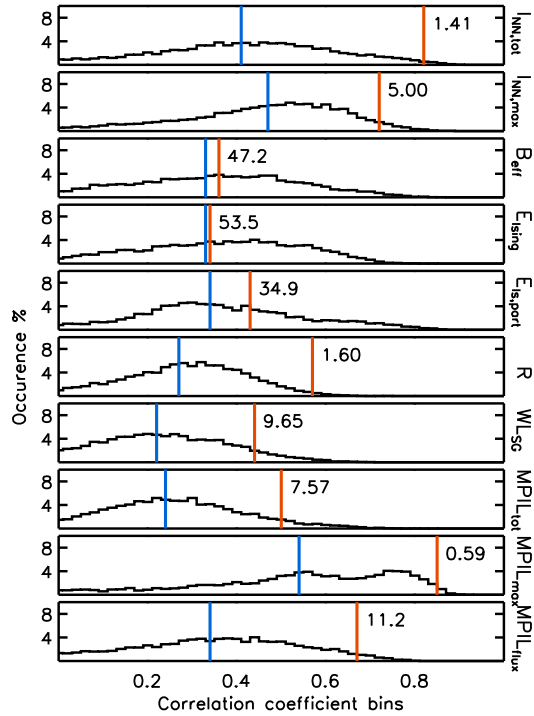}}
 \caption{Histograms of correlation coefficients between the average value of the parameters and the CME linear speed, calculated for 10,000 random combinations of 14 events out of the 32 events of the entire sample. \textit{Thick vertical lines} mark the corresponding correlations found for the fast events (\textit{red}) and, for comparison, we have also added the correlation coefficient for the entire sample (\textit{blue}). Numbers in each panel indicate the percentage of combinations with correlations higher than that for the fast events.}
\label{fig:random_correl}
 \end{figure}

Aiming to further verify that this increase of correlation observed for the 14 fast events is not a random occurrence, we examine the correlation between 10,000 random combinations of 14 out of 32. For each of these combinations we calculate the correlations between the MPIL-related properties and the CME speed (Figure~\ref{fig:random_correl}). For the length of the main MPIL there is a less than 0.6\,\% likelihood that the inferred correlation for the fast events is a random effect while for $I\sous{NN,tot}$ and $logR$ this likelihood is almost equally low ($<$1.6\,\%). For the properties that exhibit a lower overall correlation with the CME linear speed, this likelihood increases significantly. This test is another way to show that, at least for the two parameters most strongly correlated with the CME kinematic characteristics ($I\sous{NN,tot}$ and $MPIL\sous{max}$), the results presented so far are statistically significant.  

\begin{figure}
 \centerline{\includegraphics[width=1\textwidth]{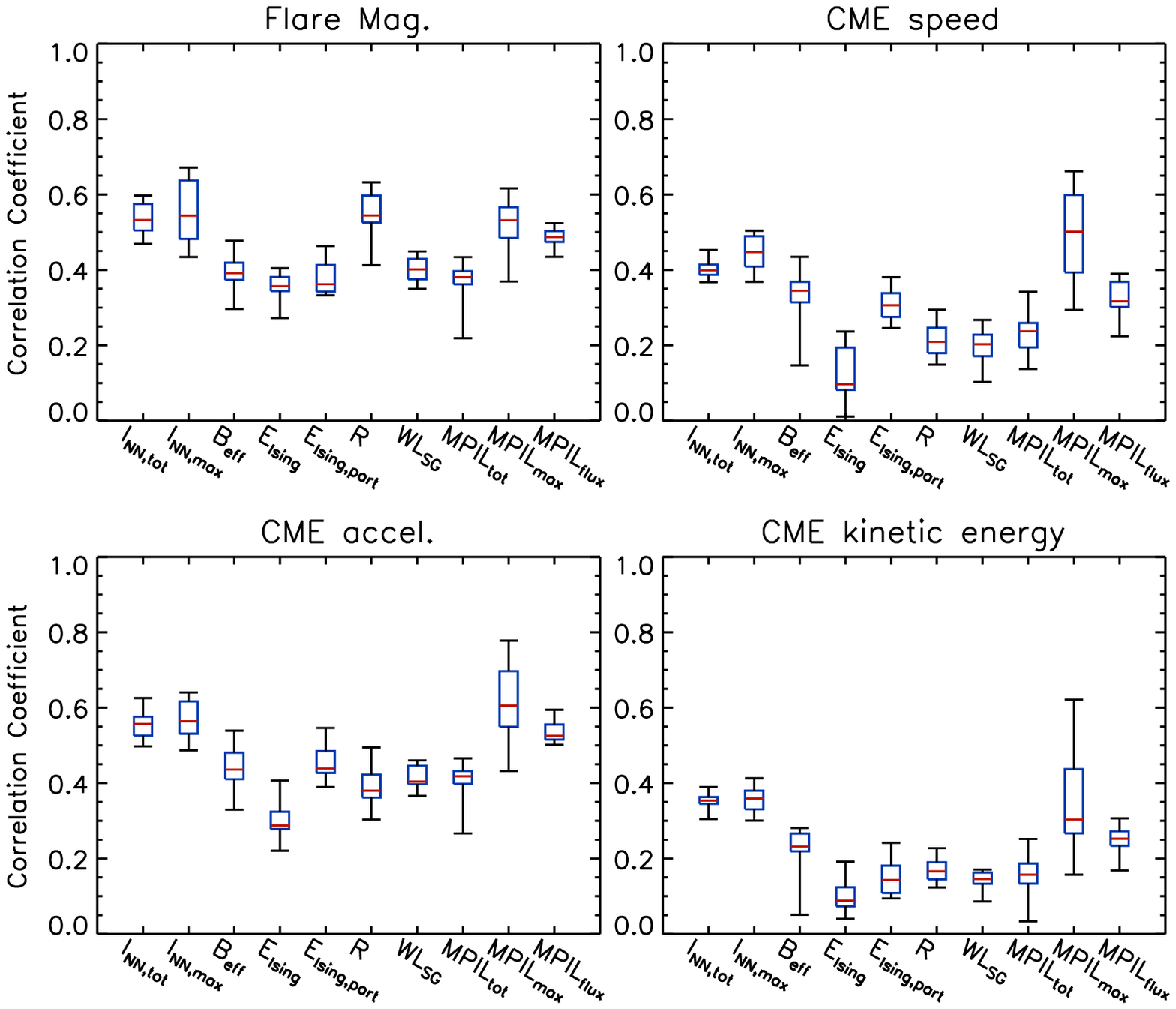}}
 \caption{Box-and-whisker plots of linear correlation coefficients calculated between every parameter value in the 24-hour window preceding the eruptive flares and eruption characteristics (see main text). For the CME acceleration, the absolute values of the correlation coefficients are plotted.}
\label{fig:box_all}
 \end{figure}

\begin{figure}
 \centerline{\includegraphics[width=1\textwidth]{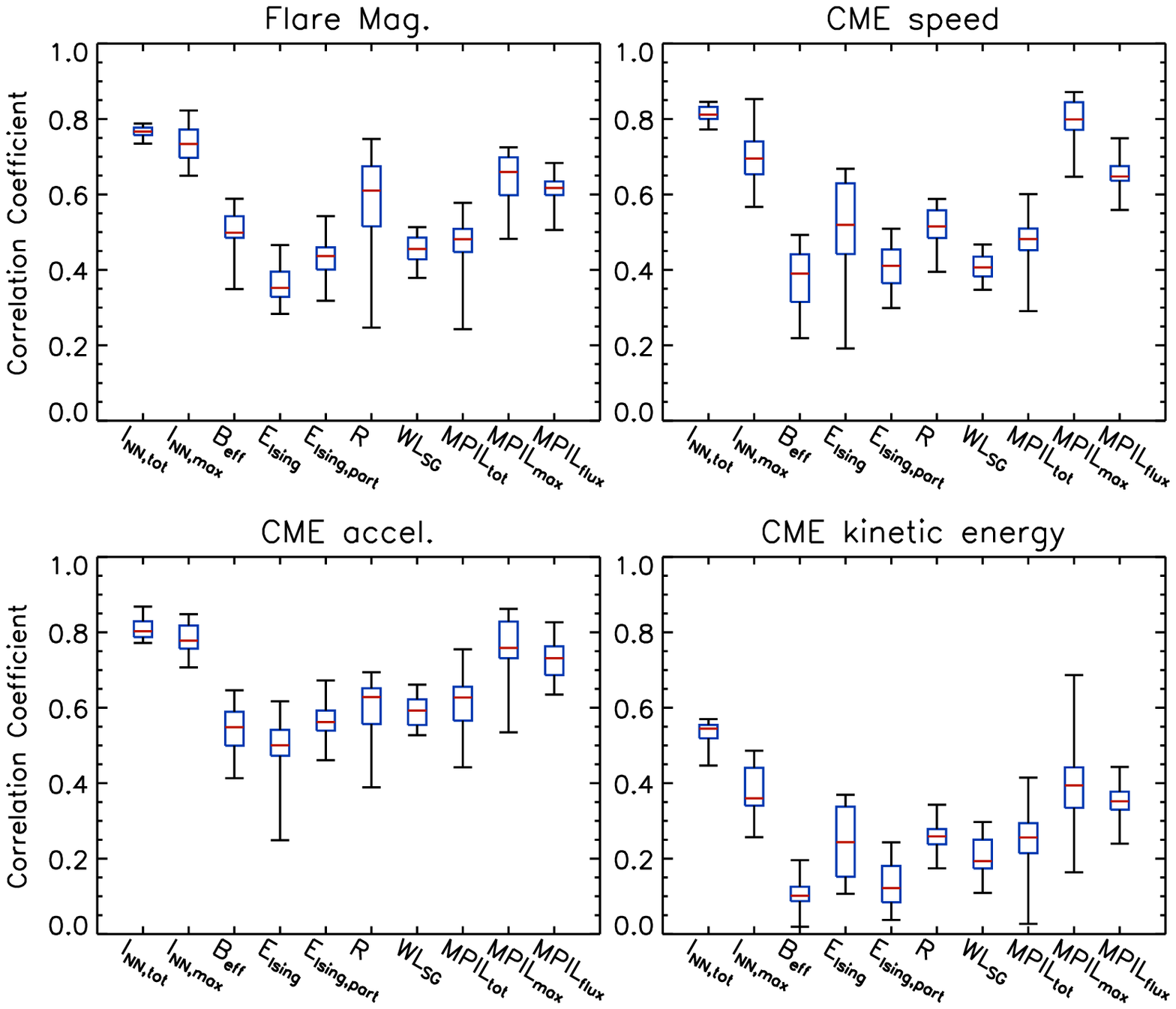}}
  \caption{Box-and-whisker plots of linear correlation coefficients calculated between every parameter value in the 24-hour window preceding the eruptive flares and eruption characteristics for fast CMEs ($> 750\,kms\sur{-1}$).}
\label{fig:box_fast}
 \end{figure}
 
\subsection{MPIL-Related Parameter Time Series and CME Kinematic Characteristics}

In this section we attempt to incorporate the temporal information of the calculated time series of properties by examining the temporal variation of the correlation of each property with the characteristics of the eruption. To do so, for each hour of the day preceding the eruptive flare we correlate the property values, instead of the 24-hour-averaged value. This results in a time series of correlation coefficients, which we plot in the box-and-whisker plots of Figures~\ref{fig:box_all} and \ref{fig:box_fast}: The boundaries of the boxes represent the first and third quartile, internal horizontal lines mark the median value; whiskers mark the minimum and maximum correlation within the time series.

These correlations are consistent with the ones presented in Figures~\ref{fig:fmag_bar} and \ref{fig:cme_bar}. In most cases, the correlation of the parameter at certain times within the 24-hour window for the flare magnitude and CME kinematic characteristics can be higher than the one exhibited by the time-averaged values. The same holds invariably both for fast CMEs and for the entire sample, although we find again that correlations are clearly stronger for fast CMEs. For example, the value of $B\sous{eff}$ is highly correlated ($\approx$0.65) with the acceleration of fast CMEs some 10-hours before the event, while $I\sous{NN,tot}$ exhibits a very small range of correlations, peaking precisely at the time of the event. This small variation may be attributed to the small variability exhibited during the day before the event, as shown in Figures~\ref{fig:time_series} and \ref{fig:rsd}. This again reminds us of the different temporal behavior exhibited by different MPIL-related properties as active regions evolve toward eruptions.

\subsection{MPIL-Related Parameters and Positional Dependence}

\begin{figure}
 \centerline{\includegraphics[width=1\textwidth]{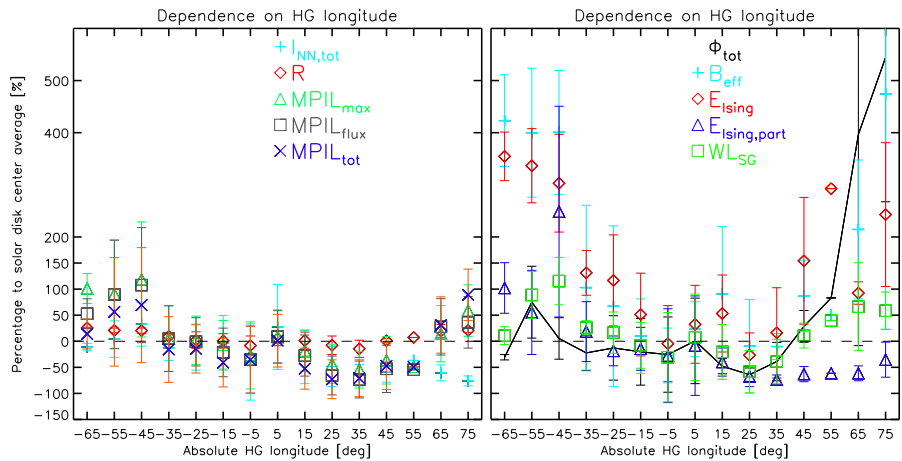}}
 \caption{Dependence of the MPIL-related parameters on heliographic (HG) longitude. The dependences are demonstrated as percentage differences from their solar disk-center values, calculated in 10$^\circ$ bins centred on each position. Error bars correspond to 1\,$\sigma$ from the average value within each bin.}
\label{fig:positions}
 \end{figure}

In previous sections we explored the different association of MPIL-related parameters with the eruption characteristics. Here we examine whether these differences can be attributed to instrumental effects since it has been clearly shown that active-region parameters calculated from magnetograms are susceptible to effects that depend on the position on the solar disk in a non-trivial way \citep{guerra18,kontogiannis18}. This is illustrated for our sample in Figure~\ref{fig:positions}, where we have also included for comparison the total unsigned magnetic flux calculated from the radial component of the magnetic field. This figure was constructed by treating all points of the 32 time series of our AR sample as 691 point-in-time observations, corresponding to as many independent heliographic (HG) locations. Our sample does not contain active regions very close to the limb, spanning over a heliographic longitude range between $-62^\circ$ and $+78^\circ$. This range was divided in $10^\circ$ bins and for each bin we calculated the average value and the standard deviation (represented by error bars). The average value was then divided by the corresponding average at the centre of the solar disk. This figure is similar to Figure~6 of \citet{kontogiannis18}.

Although the sample is significantly smaller (691 data points instead of 9454) and predictors here are calculated from the $B_{r}$-component (instead of $B\sous{LOS}$), the same predictors in the two studies show similar dependence on heliographic position. Towards the limb, all parameters exhibit deviations with respect to their disk-center-average values. The most notable deviations are found for $B\sous{eff}$ and the Ising energies apparently due to the appearance of fake polarities at large heliographic longitudes. Geometrical foreshortening of active regions toward the limb results in decreased separation distances between opposite polarity pixels and partitions and, consequently, increased values for some predictors. Additionally, it appears that $B\sous{eff}$ also ``inherits'' the East--West asymmetry detected in the $B_{r}$-component. This asymmetry, as also shown by \citet{kontogiannis18}, does not affect the non-neutralized currents, which are calculated from the derivatives of the horizontal components of the magnetic field.

It could be stated that the predictors exhibiting the best correlation with the CME and flare characteristics are indeed the ones least affected by effects relevant to the heliographic position ($R$, $WL\sous{SG}$, $E\sous{Ising}$, $E\sous{Ising,part}$, $I\sous{NN,tot}$ and $MPIL\sous{max}$). However, based on the findings of Figure~\ref{fig:positions} we cannot attribute their differences solely to these effects. For instance, $R$ appears fairly constant with heliographic longitude, so its lower correlation with CME characteristics cannot be explained solely on the grounds of positional dependence, while $MPIL\sous{tot}$, $MPIL\sous{max}$, and $MPIL\sous{flux}$ exhibit similar dependence on solar disk position, albeit noticeably different correlation strength with CME characteristics. $I\sous{NN,tot}$ and $MPIL\sous{max}$ exhibit significantly (and similarly) stronger correlation with the kinematic characteristics of CMEs, than $R$, $MPIL\sous{flux}$, and $MPIL\sous{tot}$ that permits us to conclude these two properties quantify information on the complexity and strength of MPILs in the most efficient way. This finding aligns with the exclusive cause-and-effect relation between non-neutralized currents and strong MPILs discussed in the literature \citep{georgoulis12a,georgoulis18}, even though the corresponding proxies (\textit{i.e.} $I\sous{NN,tot}$ and $MPIL\sous{max}$) are affected by projection and instrumental effects in different ways.

\section{Discussion and Conclusions}
\label{s:conclusions}

We presented a comparative analysis on the association between eruptive-flare characteristics and ten predictors of flaring activity, all characterizing the MPILs in active regions. These predictors quantify the strength, the length, the magnetic flux in the MPIL vicinity, as well as the associated net electric currents injected into the corona. Our set of predictors included not just well-established parameters but also new and promising ones that are used for the first time in automated flare prediction in the context of FLARECAST. We extended the results presented by \citet{kontogiannis17}, \citet{guerra18}, and \citet{kontogiannis18} by examining for the first time the association of these new predictors with CME characteristics. To do so, we used a small sample of unambiguously associated CMEs and source active regions, combining information from online databases widely used in heliophysics research.

Our results justify the use of these MPIL-related properties as proxies of the eruptive potential for two reasons: i) for eruptive active regions these properties obtain consistently their highest possible values during the 24 hours that precede eruptions and ii) they exhibit positive correlation with the eruption characteristics. This correlation increases noticeably for the subset of fast CMEs, a finding which appears statistically significant for the most promising predictors. The MPIL-related properties behave differently in time during the day before the eruption (Figure~\ref{fig:rsd}) and show different response to measurement errors and projection effects (Figures~\ref{fig:errors} and \ref{fig:positions}). However, it was shown that differing correlations found between various predictors and CME characteristics cannot be solely attributed to instrumental or projection effects but these differences reflect the fact that each of these MPIL-related parameters quantifies differently (and different) MPIL characteristics.

This line of reasoning is also justified by another finding. Groups of properties that represent similar aspects of MPILs show consistent behaviour. Thus, $R$ and $MPIL\sous{flux}$, both expressing the magnetic flux at the vicinity of MPILs, exhibit similarities in terms of correlations with CME properties. It is possible that these specific parameters are related to the reconnection flux, which is usually measured using the flare ribbons or the filament channel as tracers \citep{qiu05,chen06}. Similarly, $MPIL\sous{tot}$--$WL\sous{SG}$ are associated with the length of the MPILs and $B\sous{eff}$--Ising energies imply some form of connectivity between opposite polarity elements, either rudimentary (considering all possible connections) or optimized. By comparing these different groups of predictors, it appears that a predictor shows better prospects with CME characteristics when it is more closely associated with the main MPIL. This was also shown by \citet{vasantharaju18}, where manually locating the MPIL (instead of automatically) led to better correlations with the CME speed and flare magnitude.

In the same context, the two predictors that stand out in terms of correlation with eruption characteristics in this specific sample are the total unsigned non-neutralized electric currents and the length of the main MPIL. Non-neutralized electric currents require the extra information of the horizontal components of the magnetic field, and it appears \textit{a posteriori} that this information cannot be compensated for by properties relying only on the vertical (or LOS) component, with the exception of $MPIL_{max}$. Nevertheless, it is interesting how a simple measure such as the length of the main MPIL appears effective in terms of relevance to eruption characteristics. Undoubtedly, however, this result is important because it corroborates the known cause--effect relationship between strong MPILs and non-neutralized electric currents. Very recently, in a review by \citet{georgoulis19}, this relationship, starting from the formation of a strong (\textit{i.e.} sheared) MPIL, was viewed to dictate an irreversible evolution of strong-MPIL active regions toward at least one eruptive flare.  As originally explained by both \citet{georgoulis12a} and \citet{georgoulis18}, non-neutralized electric currents play an instrumental role in this evolution, facilitating the efficient accumulation of magnetic free energy and magnetic helicity around MPILs via the magnetic Lorentz force.
 
Although the focus of our study is the association of source active region with CMEs, our findings are also related to another long-standing question: are there two distinct populations of CMEs? This question was addressed by studying the properties of CMEs through coronagraphic observations. \citet{sheeley99} report two types of CMEs, impulsive and gradual, but \citet{vrsnak05} dismiss this distinction by stating that a continuum of events constitutes the CME property distributions. It should be noted, however, that the association, or lack thereof, of CMEs with flares has been central in addressing this question in many previous studies. Here we investigate only eruptive flares, and therefore all the CMEs of our sample are flare-associated. Furthermore, even though we use the distinction between fast and slow CMEs, our focus is on the source region properties. So the question must be viewed from a different angle in this study: are the active-region sources of flare-associated fast and slow CMEs the same? Or, are the precursors or specifics of the initiation of all CMEs the same? 

The variety of magnetic configurations of progenitor and ambient magnetic fields that may lead to an eruption justifies the fact that a continuum of properties is indeed observed but, in this picture, also a series of different precursors or triggering mechanisms may be in action \citep[see][for a detailed review]{chen11}. For instance, it has been found that different types of helicity evolution in source regions may lead to different types of CMEs in terms of speed \citep{park12}. Furthermore, large amounts of free energy stored in the magnetic field do not guarantee the occurrence of major events. Also dictated by the fundamental conservation of magnetic helicity, the excess energy may be channelled to a series of smaller events. These events may exhibit diverse characteristics, which may or may not be related to the main MPIL \citep[see, \textit{e.g.},][]{schrijver16}. Sophisticated MHD simulations have shown that consecutive eruptive events may take place during the emergence and reconfiguration of the magnetic field, with different mechanisms prevailing in each event as the active region evolves \citep{syntelis17}. Perhaps, then, fast CMEs exhibit a higher degree of association with non-potentiality measures because they are major events associated with a specific initiation mechanism. These events are closer to the maximum eruption scales that a given MPIL can produce at a given time. Both $I\sous{NN,tot}$ and $MPIL\sous{max}$ refer to this maximum eruptive capacity of MPILs. A slower event and a generally less energetic flare can be given by an array of MPIL values, from the strongest possible to weaker ones, hence the weaker correlations observed. 

Based on these remarks, parametric studies of active regions as they evolve towards eruptions may prove very useful to associate the evolution of certain parameters with specific mechanisms. Additionally, further investigations combining diverse sets of parameters of events and source regions \citep[as, \textit{e.g.}, in][]{murray18} and the exploitation of sophisticated tools and databases such as those of FLARECAST will provide invaluable insight into the physics of these phenomena and will greatly improve our ability to predict them.

 \begin{acknowledgments}
\noindent We would like to thank the anonymous reviewer for providing constructive comments, which improved the content of this manuscript. Also, Dr. Costas Florios for useful advice regarding the statistical methods used. This research has been funded by the European Union's Horizon 2020 research and innovation programme through the Flare Likelihood And Region Eruption foreCASTing" (FLARECAST) project, under grant agreement No. 640216 and supported by grant DE 787/5-1 of the Deutsche Forschungsgemeinschaft (DFG). The data used are courtesy of NASA/SDO, the HMI science team and the \textit{Geostationary Satellite
System} (GOES) team. This CME catalog is generated and maintained at the CDAW Data Center by NASA and The Catholic University of America in cooperation with the Naval Research Laboratory. SOHO is a project of international cooperation between ESA and NASA.
\end{acknowledgments}
\\

\noindent \textbf{Disclosure of Potential Conflicts of Interest} The authors declare that they have no conflict of interest.



 \appendix
\section{CME linear speed theshold selection}
\label{app}
\begin{figure}
 \centerline{\includegraphics[width=1\textwidth]{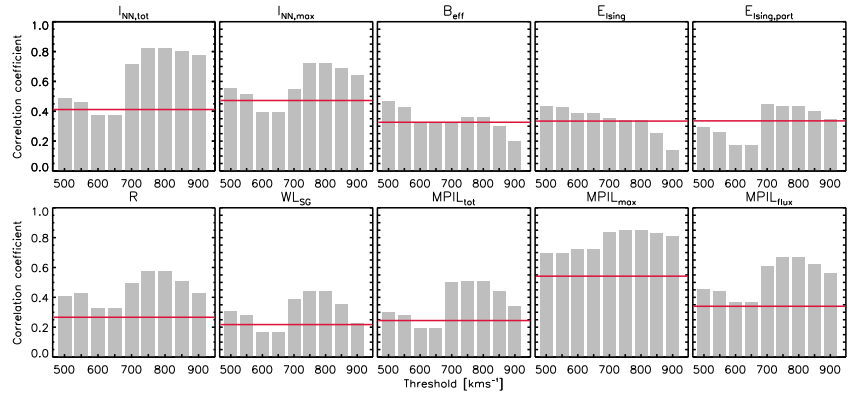}}
 \caption{Correlation coefficients between the CME linear speed and the 24-hour MPIL properties for different CME linear speed thresholds. The \textit{horizontal red lines} mark the corresponding correlation coefficients for a threshold equal to zero (\textit{i.e.} considering all values of the sample.}
\label{fig:thres}
 \end{figure}

In this analysis we examined the different correlations exhibited between eruptive flare characteristics and MPIL properties for fast and slow events. The speed threshold chosen to distinguish between these two classes was 750\,kms\sur{-1}, based on the conclusions of the statistical study of \citet{sheeley99}. As such, 750\,kms\sur{-1} is a threshold in a statistical sense, meaning that there is a degree of confluence between the two groups around this value.
In order to cross-check the validity of this threshold, we examine here the correlation between the 24-hour averaged MPIL property values with the CME linear speed, for various threshold values ranging from 500 to 900\,kms\sur{-1}. The results for all parameters are shown in Figure~\ref{fig:thres}. All parameters (with the exception of $E\sous{Ising}$) exhibit a systematic increase in the correlation coefficient between 700 and 800\,kms\sur{-1} after which the correlation coefficient drops and the size of the sample decreases. For most parameters, the maximum correlation is found within this range. Therefore, we deem that 750\,kms\sur{-1} is a reasonable threshold choice that also ensures that the two populations have comparable sizes (14 \textit{vs} 18). Additionally, Figure~\ref{fig:thres} demonstrates that any threshold selection above 700\,kms\sur{-1} would not alter significantly the conclusions of the study. 

\bibliographystyle{spr-mp-sola}
\bibliography{references}

\end{article} 

\end{document}